\documentclass[a4paper,10pt,3p,preprint,two column,sort&compress]{elsarticle}
\usepackage{graphicx}
\usepackage{setstack}
\usepackage{color}
\usepackage{natbib}
\usepackage[hyphens]{url}
\usepackage{lipsum}
\usepackage{geometry}
\usepackage[labelfont=bf]{caption}

\journal{solid state communication}

\begin{document}

\begin{frontmatter}
\title{Noise cross-correlation and Cooper pair splitting efficiency in
multi-teminal superconductor junctions. }

\author[1,2]{J.A. Celis Gil}
\author[2,3]{ S. Gomez P.}
\author[2]{William J. Herrera}
\address[1]{Kavli Institute of Nanoscience, Delft University of Technology,
Delft, The Netherlands}
\address[2]{Departamento de F\'{\i}sica,
Universidad Nacional de Colombia, Bogot\'a, Colombia}
\address[3]{Departamento de F\'{\i}sica, Universidad El Bosque, Bogot\'a,
Colombia}

\begin{abstract}
We analyze the non-local shot noise in a multi-terminal junction formed by two Normal metal leads connected to one superconductor. Using the cross Fano factor and the shot noise, we calculate the efficiency of the Cooper pair splitting. The method is applied to $d$-wave and iron based superconductors. We determine that the contributions to the noise cross-correlation are due to crossed Andreev reflections (CAR), elastic cotunneling, quasiparticles transmission and local Andreev reflections. In the tunneling limit, the CAR contribute positively to the noise cross-correlation whereas the other processes contribute negatively. Depending on the pair potential symmetry, the CAR are the dominant processes, giving as a result a high efficiency for Cooper pair split. We propose the use of the Fano factor to test the efficiency of a Cooper pair splitter device.
\end{abstract}

\ead{jherreraw@unal.edu.co}

  \end{frontmatter}

\section{ \label{SEC:Introduction} Introduction}

Entanglement states between photons have been very well developed \cite%
{Salter}. However, entanglement between electrons in solid system is
difficult to create because the electrons are immersed in a macroscopic
ground state, which prevents the straightforward generation of entangled
pairs of electrons.

The use of a superconductor to produce entangled electrons has been proposed
\cite%
{Rycerz,Chtchelkatchev,Herrmann,Martin1999,Zhang2009,Lesovik2001,Recher2001}
because its ground state is composed of Cooper pairs. A Cooper pair in the
superconductor can be break up into two nonlocal entangled electrons that
enter into different normal metal leads via the Cooper pair splitting (CPS)
\cite{Byers1995,Deutscher2000,Burset201402,Wei-Jiang,Zuo201417}. The CPS has
been studied theoretically and experimentally \cite%
{Hofstetter,Wei,Braunecker136806,Soller2013,Williamtubo}, opening a door to
test Bell inequalities in the solid state \cite{Kawabata,Forgues2015}.

The Cooper pair splitting process is analog to CAR, where an incoming
electron from one of the leads is reflected as a hole in the other one,
inducing a Cooper pair in the superconductor. Typically, to analyze the CAR,
the electrical current has been used to test the Bell inequalities; it is
important to analyze not only the currents through every lead, but also the
correlations between them, which can be determined through the noise.

According to its origin, the noise is classified into two types: one due to
thermal fluctuations, which is known as Nyquist-Johnson noise \cite{Blanter}%
, and the other due to the discrete behavior of the electric charge, known
as shot noise \cite{Beenakker}. It is possible to obtain with shot noise
measurements information not commonly obtained with conductance
measurements. In high $T_{C}$ superconductors, the shot noise has
been studied for plain junctions where shot noise is affect by the pair
potential symmetry \cite{Zhu1999, Tanaka2000}.

The nonlocal shot noise or noise cross-correlation between two electrodes
connected to a superconductor reveals a change of the sign in the
correlation between currents \cite{Texier,Chevallier,Faoro}. For example,
for two electrodes connected to a normal metal, the nonlocal shot noise
shows a negative crossed correlation, which indicates that when the
electrical current in one lead increases, the current in the other lead
decreases \cite{Mishchenko,KieBlich,Qin2008}. However the nonlocal shot
noise for two leads connected to a superconductor can exhibits positive
values \cite{Agrait,Melin2} that is, the electrical current in the two leads
could increase or decrease at the same time.

The noise cross-correlation has been studied for two quantum point leads
connected to a superconductor \cite%
{Melin1,Samuelsson,Samuelsson2,Wu2014,Chen2015,Braggio2011155};
nevertheless, the nonlocal shot noise has not been studied for two
electrodes connected to $d$-wave or iron based superconductors, where the
pair potential symmetry can affect the transport properties of the system.
In the present paper we show an analytic approach that allows us to find the
current-current correlation and separate the contributions due to the
different processes. In particular we analyze the Cooper pair split
contribution. For this, we use the Hamiltonian approach and the
non-equilibrium Green functions in Keldysh formalism. We consider typical
pair potential symmetries for cuprates and iron based superconductors which
means $d$, $s_{++}$ and $s_{+-}$ symmetries \cite%
{Kirtley,Yong,Biswas,Emamipour201617,Jia20122133,Wei2010777,Pradhan201653,Zhao2012660}%
. In addition, we consider a $s$-wave compound added to the $d$-wave
superconductors, thus we study how the magnitude and phase of the pair
potential affect the noise cross-correlation. We analyze the symmetric case
when the two leads are connected to the same voltage ($V_{a}=V_{b}=V$) and
the non-symmetric case when the two leads are connected to a voltage
difference $V$ ($V_{a}=0,V_{b}=V$). We find that the positive noise
cross-correlation is favored by symmetric applied voltages.

\section{Shot noise cross-correlation}

\begin{figure}[h]
\includegraphics[width=6.0cm]{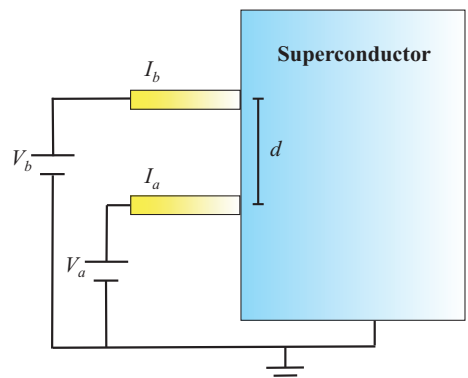}
\caption{Dagram of two electrodes $a(b)$ separated by a distance $d$ and
connected to a superconductor. The superconductor is grounded and the leads
are at voltages $V_{a}$ and $V_{b}$ respectively.}
\label{Fig.1}
\end{figure}

\begin{figure}[tbp]
\includegraphics[width=6.0cm]{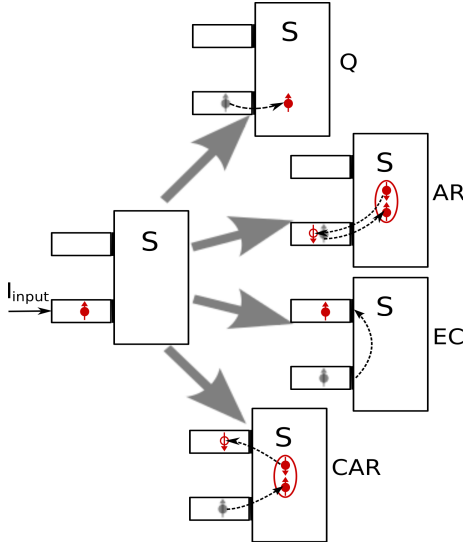}
\caption{Diagram of the different processes considered. An incoming electron
from a lead could be: transmitted as a quasiparticle (Q); reflected as a
hole in the same lead inducing a Cooper pair in the superconductor (AR);
reflected as an electron in the other lead (EC); or reflected as a hole in
the other lead inducing a Cooper pair in the superconductor (CAR).}
\label{Fig.2}
\end{figure}

The system considered is formed by two one-dimensional normal metal leads
connected to a semi-infinite superconducting region Fig. (\ref{Fig.1}). The
lead $a(b)$ is connected to voltage $V_{a(b)},$ while the superconductor is
grounded. It has been demonstrated that there are two processes that
contribute to the cross conductance; the CAR and the elastic cotunneling
(EC) \cite{Williamdos}, see Fig. \ref{Fig.2}. When the $V_{a}=0$ and $%
V_{b}=V $ the nonlocal differential conductance is given by

\begin{equation}
\sigma _{ab}=\frac{dI_{a}}{dV_{b}}=\frac{2e^{2}}{h}\left(
T_{CAR}-T_{EC}\!\right) ,  \label{Conductance}
\end{equation}%
where $T_{CAR}$ and $T_{EC}$ are the transmission coefficients of the CAR
and EC respectively and that can be written in terms of the Green function as

\begin{eqnarray}
T_{CAR} &=&4t^{4}|\check{G}_{ab,eh}^{r}(E)|^{2}, \\
T_{EC} &=&4t^{4}|\check{G}_{ab,ee}^{r}(E)|^{2},
\end{eqnarray}%
where $t$ is the hopping parameter coupling the leads and the
superconductor, $\check{G}_{ab}^{r}(E)$ is the nonlocal green function in
the superconducting region between $a$ and $b$, the subindex $ee(h)$ denotes
the electron-electron (hole) component. The contributions of $T_{CAR}$ and $%
T_{EC}$ processes decrease when the distance between the leads increases and
depend on the symmetry of the pair potential \cite{Williamdos}.

\begin{figure}[tbp]
\includegraphics[width=6.0cm]{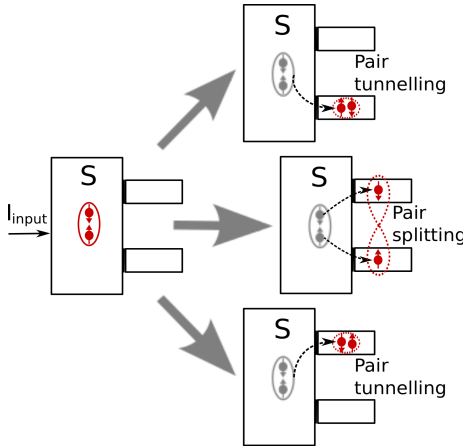}
\caption{Basic transport processes in a Cooper pair splitter: the electrons
of a Cooper pair either leave the superconductor into the same arm (pair
tunneling) or split up into different arms (pair splitting).}
\label{Fig.3}
\end{figure}

Our aim is to obtain the noise cross-correlations and the Fano factor. For $%
s $-wave superconductors, the nonlocal shot noise has been calculated \cite%
{Melin2}, showing that, whereas the CAR contribute positively to the crossed
correlation between the currents in the two leads \cite%
{Russo,Beckmann,Anindya}, the EC contributes negatively \cite{Cadden}. Local
processes like the Andreev reflections (AR) and quasiparticles transmission
(Q) contribute negatively to the nonlocal shot noise (see Fig. \ref{Fig.2}).
Whereas the AR are equivalent to Cooper pair tunneling in one lead, the CAR
are equivalent to CPS \cite{Schindele, Hofstetter2}; hence, we are
interested in analyzing the positive contributions to the nonlocal shot
noise (see Fig. \ref{Fig.3}).

We use the Hamiltonian approach to study this system and the Green
functions, and Keldysh formalism \cite{Cuevas} in order to find the nonlocal
shot noise between the leads $\beta $ and $\beta ^{\prime }$ at frequency $%
\omega $ (for details see appendix A),

\begin{eqnarray}
S_{\beta \beta ^{\prime }}\left( \omega \right) &=&\frac{2e^{2}t^{4}}{h}\int
dE\left[ K_{\beta \beta ^{\prime }}(E,\varepsilon )\right. \\
&&\left. +K_{\beta ^{\prime }\beta }(\varepsilon ,E)\right] ,  \nonumber
\end{eqnarray}%
where

\[
\varepsilon =E+\hbar \omega .
\]

The kernel $K_{\beta \beta ^{\prime }}\left( E,\varepsilon \right)$ can be
rewritten as

\begin{eqnarray}
K_{\beta \beta ^{\prime }}\left( E,\varepsilon \right) &=&K_{\beta \beta
^{\prime }}^{1}(E,\varepsilon )+K_{\beta \beta ^{\prime }}^{2}(E,\varepsilon
) \\
&&+K_{\beta \beta ^{\prime }}^{3}(E,\varepsilon )+K_{\beta \beta ^{\prime
}}^{4}(E,\varepsilon ),  \nonumber
\end{eqnarray}%
where

\begin{eqnarray}
K_{\beta \beta ^{\prime }}^{1}(E,\varepsilon ) &=&4\pi ^{2}(1-f(\varepsilon
))\mathrm{Tr}\left[ \check{N}_{\beta }\check{A}_{L}(E)\right.  \nonumber \\
&&\check{f}_{L}(E)\check{\rho}_{L}(E)\check{q}_{\beta ^{\prime
}}^{1}(E,\varepsilon )\check{\rho}_{R}(\varepsilon )  \nonumber \\
&&\left. \check{A}_{R}^{\dagger }(\varepsilon )\right] ,  \nonumber \\
K_{\beta \beta ^{\prime }}^{2}(E,\varepsilon ) &=&4\pi
^{2}f(E)(1-f(\varepsilon ))\mathrm{Tr}\left[ \check{N}_{\beta }\right.
\nonumber \\
&&\check{G}_{LL}^{r}(E)\check{p}\check{\rho}_{R}(E)\check{q}_{\beta ^{\prime
}}^{2}(E,\varepsilon )  \nonumber \\
&&\left. \check{\rho}_{R}(\varepsilon )\check{A}_{R}^{\dagger }(\varepsilon )%
\right] ,  \nonumber \\
K_{\beta \beta ^{\prime }}^{3}(E,\varepsilon ) &=&4\pi ^{2}\mathrm{Tr}\left[
\check{N}_{\beta }\check{A}_{L}(E)\check{f}_{L}(E)\right.  \nonumber \\
&&\check{\rho}_{L}(E)\check{q}_{\beta ^{\prime }}^{3}(E,\varepsilon )(1-%
\check{f}_{L}(\varepsilon ))  \nonumber \\
&&\left. \check{\rho}_{L}(E)\check{p}\check{G}_{RR}^{a}(\varepsilon )\right]
,  \label{kernels} \\
K_{\beta \beta ^{\prime }}^{4}(E,\varepsilon ) &=&4\pi ^{2}f(E)\mathrm{Tr}%
\left[ \check{N}_{\beta }\check{G}_{LL}^{r}(E)\check{p}\right.  \nonumber \\
&&\check{\rho}_{R}(E)\check{q}_{\beta ^{\prime }}^{4}(E,\varepsilon )(1-%
\check{f}_{L}(\varepsilon ))  \nonumber \\
&&\left. \check{\rho}_{L}(E)\check{p}\check{G}_{RR}^{a}(\varepsilon )\right]
,  \nonumber
\end{eqnarray}%
with

\begin{eqnarray}
\check{q}_{\beta ^{\prime }}^{1}(E,\varepsilon ) &=&\check{A}_{L}^{\dagger
}(E)\check{N}_{\beta ^{\prime }}\check{A}_{R}(\varepsilon )-\check{p} \\
&&\check{G}_{RR}^{a}(E)\check{N}_{\beta ^{\prime }}\check{G}%
_{LL}^{r}(\varepsilon )\check{p},  \nonumber \\
\check{q}_{\beta ^{\prime }}^{2}(E,\varepsilon ) &=&\check{p}\check{G}%
_{LL}^{a}(E)\check{N}_{\beta ^{\prime }}\check{A}_{R}(\varepsilon ) \\
&&-\check{A}_{R}^{\dagger }(E)\check{N}_{\beta ^{\prime }}\check{G}%
_{LL}^{r}(\varepsilon )\check{p},  \nonumber \\
\check{q}_{\beta ^{\prime }}^{3}(E,\varepsilon ) &=&\check{A}_{L}^{\dagger
}(E)\check{N}_{\beta ^{\prime }}\check{G}_{RR}^{r}(\varepsilon )\check{p} \\
&&-\check{p}\check{G}_{RR}^{a}(E)\check{N}_{\beta ^{\prime }}\check{A}%
_{L}(\varepsilon ),  \nonumber \\
\check{q}_{\beta ^{\prime }}^{4}(\!E,\varepsilon ) &=&\check{p}\check{G}%
_{LL}^{a}(E)\check{N}_{\beta ^{\prime }}\check{G}_{RR}^{r}(E)\check{p} \\
&&-\check{A}_{R}^{\dagger }(E)\check{N}_{\beta ^{\prime }}\check{A}%
_{L}(\varepsilon ),  \nonumber \\
\check{A}_{L}(E) &=&\check{I}+\check{G}_{LR}^{r}(E)\check{p}, \\
\check{A}_{R}(E) &=&\check{I}+\check{G}_{RL}^{r}(E)\check{p}.
\end{eqnarray}

In this equations, $\check{G}_{ij^{\prime }}^{r}$ is the retarded Green
function between the region $i$ and $j$ in Nambu $\otimes $ electrodes
space, the subindex $L$ denotes the "left" region of leads and the subindex $%
R$ the right superconducting region, $\check{\rho}_{i}$ is the local density
of states, $\check{\rho}_{i}=-\mathrm{Im}(\check{g}_{ii}^{r})/\pi ,$ with $%
\check{g}_{ii}$ the unperturbed Green function, $\check{p}=t\ast \check{%
\sigma}_{z}$ is the hopping matrix, \ and

\begin{eqnarray}
\check{N}_{a} &=&\left(
\begin{array}{cc}
\hat{I} & \hat{0} \\
\hat{0} & \hat{0}%
\end{array}%
\right) ,\check{N}_{b}=\left(
\begin{array}{cc}
\hat{0} & \hat{0} \\
\hat{0} & \hat{I}%
\end{array}%
\right) ,  \nonumber \\
\check{\sigma}_{z} &=&\left(
\begin{array}{cc}
\hat{\sigma}_{z} & \hat{0} \\
\hat{0} & \hat{\sigma}_{z}%
\end{array}%
\right) ,  \label{MAtrixN} \\
\check{f}_{L} &=&\left(
\begin{array}{cc}
\hat{f}_{La} & \hat{0} \\
\hat{0} & \hat{f}_{Lb}%
\end{array}%
\right) ,  \nonumber \\
\hat{f}_{L\beta } &=&\left(
\begin{array}{cc}
f_{e\beta } & \hat{0} \\
\hat{0} & f_{h\beta }%
\end{array}%
\right) ,  \nonumber
\end{eqnarray}%
with $f_{e\beta }=f\left( E-eV_{\beta }\right) $, $f_{h\beta }=f\left(
E+eV_{\beta }\right) $ the Fermi distributions for electron and hole like
quasiparticles at the lead $\beta $. Each kernel $K_{\beta \beta ^{\prime
}}^{\alpha }$ describes a process that contributes to the electrical current
fluctuations and can be classified according to its origin (\ref{kernels}).
The kernel $K_{\beta \beta ^{\prime }}^{1}\left( E,\varepsilon \right) $ can
be interpreted as the dispersion of particles injected from the $\beta $
lead to the superconductor, where the product $\left( 1-f\left( \varepsilon
\right) \right) \check{f}_{L}\left( E\right) $ relates the electron
transmission from the left side to the superconductor. Similarly in $%
K_{\beta \beta ^{\prime }}^{4}\left( E,\varepsilon \right) $ the product $%
f\left( E\right) \left( 1-\check{f}_{L}\left( \varepsilon \right) \right) $
can be interpreted as the dispersion of particles from the superconductor to
the $\beta $ lead. The $K_{\beta \beta ^{\prime }}^{2}\left( E,\varepsilon
\right) $ term corresponds to the quasiparticles dispersion and the product $%
f(E)(1-f(\varepsilon ))$ is related with thermal fluctuations; and finally
in $K_{\beta \beta ^{\prime }}^{3}\left( E,\varepsilon \right) $ the product
$\check{f}_{L}(E)\check{\rho}_{L}(E)\check{q}_{\beta ^{\prime
}}^{3}(E,\varepsilon )(1-\check{f}_{L}(\varepsilon ))\check{\rho}_{L}(E)$
corresponds to hole-hole, electron-hole and electron-electron reflections
(further details can be found in \ref{Sec:Modeling the system}).

At zero temperature and zero frequency all the products $f\left( E\right)
\left( 1-f\left( E\right) \right) $ vanish \cite%
{Tsuchikawa2001224,PhysRevB.60.9750}, therefore the thermal fluctuations
become zero and we obtain the shot noise. For $\beta =\beta ^{\prime }$ we
obtain the local shot noise at the same lead, while for $\beta \neq \beta
^{\prime }$ we obtain the nonlocal shot noise, which we analyze in two cases
with respect to the applied voltage: the non-symmetric (n-sy) case, where
the lead $a$ is grounded and the lead $b$ is at voltage $V $; and the
symmetric case (sy), where the two leads are at the same voltage $V$.

For non-symmetric voltages and in the tunneling limit, the noise
cross-correlation expression becomes to

\begin{eqnarray}
S_{ba}^{n-sy}(V) &=&S_{CAR}(V)+S_{EC}(V),  \label{dif_shot_noise1} \\
S_{CAR}(V) &=&\frac{4e^{2}}{h}\int_{0}^{eV}dE\!\left[ T_{CAR}\left( E\right) %
\right] , \\
S_{EC}(V) &=&-\frac{4e^{2}}{h}\int_{0}^{eV}dE\!\left[ T_{EC}\left( E\right) %
\right] ,
\end{eqnarray}%
in this case $T_{CAR}$ and $T_{EC}$ are calculated from the unperturbed
Green function of the superconducting region $\check{G}_{RR}^{r}(E)=\hat{g}%
^{r}(E)$

\begin{eqnarray}
T_{CAR}\left( E\right) &=&4t^{4}|\check{g}_{ab,eh}^{r}(E)|^{2}, \\
T_{EC}\left( E\right) &=&4t^{4}|\check{g}_{ab,ee}^{r}(E)|^{2}.
\end{eqnarray}

The noises $S_{CAR}$ and $S_{EC}$ correspond to noise due to crossed
electron-hole and electron-electron reflection probabilities respectively.
While CAR contributes positively to the shot noise cross correlations, EC
contributes negatively. The differential nonlocal shot noise is defined as

\begin{eqnarray}
\frac{dS_{ba}^{n-sy}(V)}{dV} &=&\frac{4e^{3}}{h}\left( T_{CAR}\left(
eV\right) \right. \\
&&\left. -T_{EC}\left( eV\right) \right) ,  \nonumber
\end{eqnarray}%
and is proportional to $\sigma _{ab},$Eq. \ref{Conductance}, $%
dS_{ba}^{n-sy}/dV=2e\sigma _{ab}$. Therefore for low transparencies and
non-symmetric voltages, we get only CAR and EC contributions to the
differential crossed shot noise. However, for symmetric voltages in the
tunneling limit, the contributions due to EC for voltages lower than $%
\left\vert \Delta (\mathbf{k})\right\vert $ are zero, then the shot noise
can be written as

\begin{equation}
S_{ba}^{sy}(V)=\frac{8e^{2}}{h}\int_{0}^{eV}dE\left[ T_{CAR}\left( E\right) %
\right] ,  \label{sy}
\end{equation}%
and therefore the differential nonlocal shot noise is written as

\begin{equation}
\frac{dS_{ba}^{sy}(V)}{dV}=8e^{3}T_{CAR}\left( eV\right) .
\label{dif_shot_noiseV}
\end{equation}

For higher transparencies, other processes contribute to the noise
cross-correlations \cite{Golubev}. Then, for the symmetric case with
voltages smaller than $\left\vert \Delta (\mathbf{k})\right\vert $, the main
contribution to the shot noise is due to CAR. For higher voltages, there are
other processes like quasiparticles transmission and Andreev reflections
which can contribute positively or negatively to shot noise. EC contributes
negatively and appear by means of intermediate propagators. For any voltage
the nonlocal shot noise in tunnelling limit can be written as

\begin{eqnarray}
S_{ba}^{sy}(\!V\!) &=&\frac{4e^{2}}{h}\int_{0}^{eV}dE\left[ 2T_{CAR}\left(
E\right) \right.  \label{shotnoisesym2a} \\
&&-2T_{EC}\left( E\right) +\left. \delta T_{EC}\left( E\right) \right] ,
\nonumber
\end{eqnarray}%
with

\begin{eqnarray}
\delta T_{EC}\left( E\right) &=&4t^{4}\left( \hat{g}_{ba,ee}^{r}(E)\hat{g}%
_{ab,ee}^{r}(E)\right. \\
&&\left. +\hat{g}_{ab,ee}^{a}(E)\hat{g}_{ba,ee}^{a}(E)\right) .  \nonumber
\end{eqnarray}

From equation (\ref{shotnoisesym2a}) we can see that for voltages lower than
$\left\vert \Delta (\mathbf{k})\right\vert $, EC contribution cancels itself
out because the nonlocal Green functions are always real, $\delta
T_{EC}=2T_{EC}$ recovering the equation (\ref{sy}).

\subsection{The Fano factor and the efficiency}

The Fano factor $(F)$ is defined as the ratio between the shot noise and the
$2e$ multiplied by the electrical current \cite{Texier}

\begin{eqnarray}
F=\frac{S}{2eI}.
\end{eqnarray}

In the tunneling limit the Fano factor provides information about the
effective electric charge. We analyze the cross correlations using the Fano
factor in terms of the noise cross-correlation and cross current considering
the equation used by Samuelsson et. al \cite{Samuelsson}

\begin{equation}
F_{\beta \beta ^{\prime }}=\frac{S_{\beta \beta ^{\prime }}}{2e\sqrt{%
I_{\beta }I_{\beta ^{\prime }}}},
\end{equation}%
where $I_{\beta }$ is the current at the lead $\beta $. According to
equation \ref{dif_shot_noise1}, in the tunneling limit, with non-symmetric
voltages, shot noise and current contributions are due to CAR and EC

\begin{equation}
F_{ab}^{n-sy}=\frac{S_{CAR}+S_{EC}}{2e\sqrt{I_{a}I_{b}}},
\end{equation}%
and in the case of symmetric voltages and the shot noise contributions get
reduced to only CAR, for $V<|\Delta(\mathbf{k})|$, therefore the cross Fano
is

\begin{equation}
F_{ab}^{sy}=\frac{S_{CAR}}{2e\sqrt{I_{a}I_{b}}}.
\end{equation}

It is expected that the nonlocal Fano factor takes values between $-1$ and $%
1 $. The explanation arises from the effective electric charge. If we
examine the electrical current in every normal metal lead, we find that the
electrical current in every lead is due to electron or hole transmission;
thus the maximum and the minimum values that the effective electric charge
can have are $e$ and $-e$ respectively. This Fano factor sign also gives
information about sign of the noise cross-correlation. A Fano factor bigger
than zero means positive cross correlations dominance, whereas a Fano factor
smaller than zero means negative cross correlations dominance.

Considering all the processes that contribute to the nonlocal shot noise in
the tunneling limit, we define the efficiency of the device ($\eta $) as the
ratio between the differential nonlocal shot noise due to CAR and the sum of
the absolute values of the contributions due to CAR and EC (\ref%
{dif_shot_noiseV}).

\begin{equation}
\eta =\frac{S_{CAR}}{|S_{CAR}|+|S_{EC}|}.  \label{efficiency}
\end{equation}

For the symmetric case, in the tunneling limit and voltages lower than $%
\left\vert \Delta (\mathbf{k})\right\vert $ the contributions to the
nonlocal shot noise are only due to CAR; then the efficiency and the Fano
factor are equal to one. If we set higher voltages, the other processes
contribute to the noise cross-correlation, which indicates that the
efficiency and the Fano factor decrease. This result implies that the Fano
factor provides information concerning the efficiency of the device. In the
next section, we show the efficiency and the Fano factor using different
pair potential symmetries and voltages at the leads.

\section{Results}

We calculate the nonlocal shot noise as a function of the distance between
the leads and the transmission coefficient which is related to the hopping
parameter $t$ by
\begin{equation}
T_{N}=\frac{4t^{2}}{\left( 1+t^{2}\right) ^{2}}.
\end{equation}

We analyze the shot noise and the Fano factor using typical symmetries for
HTc superconductors (HTcS), with $d$, $d+is$ and, $s_{++}$ and $s_{+-}$
symmetries. In this work we have fixed $\Delta _{0}=20\ meV$ and the ratio $%
\Delta _{0}/E_F \simeq 10^{-1}$, which are typical values for a HTcS. We
consider non-isotropic symmetries for the superconductor pair potential
which depends on the wave vector. For $d$-wave superconductors the pair
potential is $\Delta (\theta )=\Delta _{0}\cos 2(\theta -\alpha )$, with $%
\alpha $ the angle between the crystallographic axes of the superconductor
and the normal direction to the surface. We consider two symmetries: the
first one is $d_{x^{2}-y^{2}}$ where $\alpha =0$ and the other one is $%
d_{xy} $ where $\alpha =\pi /4$.

We consider mixed symmetries, including a small isotropic component to the $%
d $-waves, such that the magnitude and phase of the pair potential are
affected \cite{Kirtley, Yong, Biswas}.

Finally, to describe the shot noise in iron based superconductors, we
consider two multiband models the $s_{++}$ and the $s_{+-}$. In the first
one the phase difference between the two gaps is $0$, whereas in the second
one it is $\pi $ so that $\Delta _{1}/\Delta _{2}=\pm |\Delta _{1}|/|\Delta
_{2}|$. $\Delta _{1(2)}$ is the pair potential in the band $1(2)$ \cite%
{Kamihara, Golubov, Zhang}.

\subsection{$s$-wave superconductors}

If we consider an isotropic symmetry for the superconductor pair potential $%
\Delta (\mathbf{k})=\Delta _{0}$. For non-symmetric case and in the
tunneling limit, the differential nonlocal shot noise exhibits oscillations
as a function dependent on the distance between the two leads. At this limit
the contribution to the noise cross-correlations due to Q and AR is much
smaller than CAR and EC, it meaning that CAR and EC compete between them.
For higher values of the transmission $T_{N}$, the noise cross-correlation
becomes negative due to the increase in processes that contribute negatively
to the cross-correlations such as quasiparticles transmission (see Fig. \ref%
{Fig.4}). Finally for hight transparencies $T_{N}\rightarrow 1$, the
differential nonlocal shot noise is completely positive because for a
transparent lead the CAR have additional contributions of intermediate
propagators in the superconducting region \cite{Melin2}. In addition, the
noise cross-correlation exhibits an exponential decay with respect to the
distance between the leads (see Fig. \ref{Fig.5} a)). These results are in
agreement with those obtained by R. Melin et al. \cite{Melin2}.

\begin{figure}[h]
\centering{}\includegraphics[width=6.3cm]{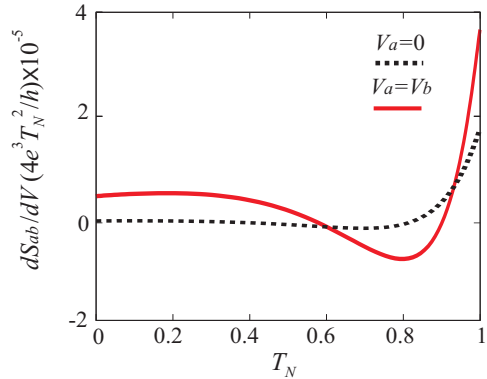}
\caption{The spatial-averaged differential shot noise cross-correlation as a
function of the transmission coefficient for a $s$ wave superconductor, at $1%
\protect\xi _{0}$ and $eV=0$, for non-symmetric voltages $V_a$=0 and
symmetric voltages $V_a = V_b$.}
\label{Fig.4}
\end{figure}

When $V_{a}=V_{b}$ at the tunneling limit, the cross correlation between the
currents is positive because EC contribution is zero and no local shot noise
is due only to CAR. Similarly to the non-symmetric case, when $T_{N}$
increases the shot noise takes negative values. This behavior occurs because
of the higher AR and Q contributions, see Fig. \ref{Fig.4}. Finally, for
high transparencies, the shot noise becomes positive due to contributions
from various processes with intermediate propagators between the two leads,
see Fig. \ref{Fig.5} b).

\begin{figure}[ht]
\includegraphics[width=6.3cm]{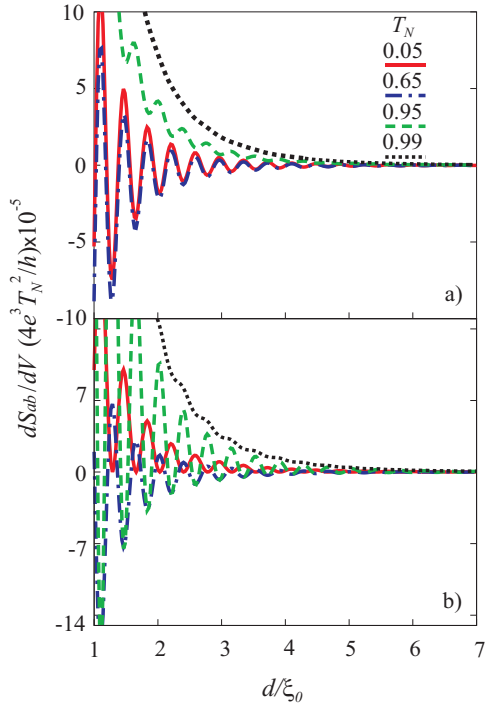}
\caption{Differential shot noise cross-correlation as a function of the
distance between the two leads for different transparencies, for a $s$ wave
superconductor, for a) non-symmetric voltages and b) symmetric voltages.
With $\protect\xi _0$ the BCS coherence length. }
\label{Fig.5}
\end{figure}

\begin{figure}[h]
\centering{}\includegraphics[width=6.3cm]{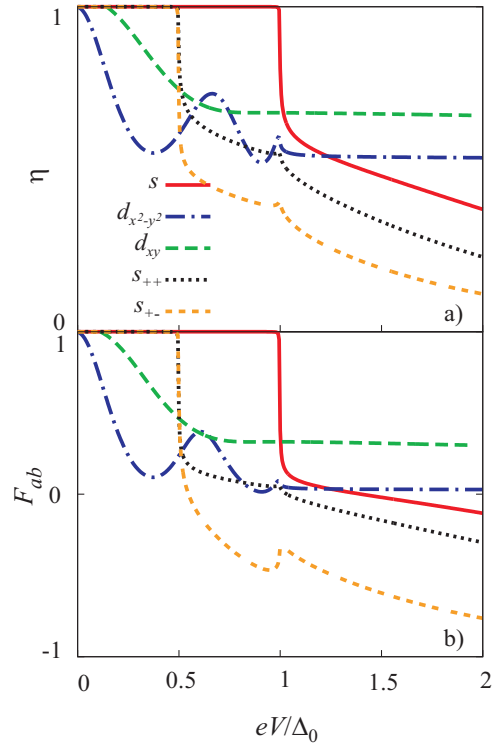}
\caption{a) Cross Andreev reflections efficiency and b) Cross Fano factor as
a function of the energy, for every symmetry considered using $V_a=V_b$ and $%
T_N=0.05$. For the symmetries $s_{+-}$ and $s_{++}$ $\protect\alpha=0.5$.}
\label{Fig.6}
\end{figure}

For symmetric voltages smaller than $\left\vert \Delta \left( \mathbf{k}%
\right) \right\vert $ and at the tunneling limit, due to the cancellation of
the EC and the negligible contributions from AR and Q, CAR become the
dominant processes giving as a result an efficiency equal to one. For higher
voltages, Q and AR contributions increase causing a decrease in the
efficiency and the Fano factor (Fig. \ref{Fig.6}).

\subsection{$d$-wave superconductors}

Unlike $s$-wave superconductors, for $d_{x^{2}-y^{2}}$ symmetry, when $%
V_{a}=0$ and $V_{b}=V$, the differential shot noise is positive for every
transmission value, indicating that this symmetry favors positive cross
correlations, in particular for low transparencies.

When we set $V_{a}=V_{b}$, the contribution to the shot noise due to CAR is
twice the obtained for non-symmetric voltages whereas the EC contribution
cancel. This is reflected in the increase of the differential cross
correlation shot noise in the tunneling limit. Similarly to $s$-wave
superconductors, the shot noise is positive for low transparencies; however,
as $T_{N}$ increases the differential shot noise becomes negative due to AR
and Q. For high transparencies, the differential shot noise becomes positive.

For this pair potential symmetry, the cross correlation oscillations do not
disappear for high transparencies. This is due to the relative phase of the
pair potential, which allows constructive and destructive interferences in
the leads. We appreciate an algebraic decay of the shot noise with respect
to the distance between the leads, proportional to $1/d^{2}$ (Fig. \ref%
{Fig.7} a) and b)), in contrast to the exponential decay of the $s$ wave
superconductors.

\begin{figure}[h]
\centering{}\includegraphics[width=7.3cm]{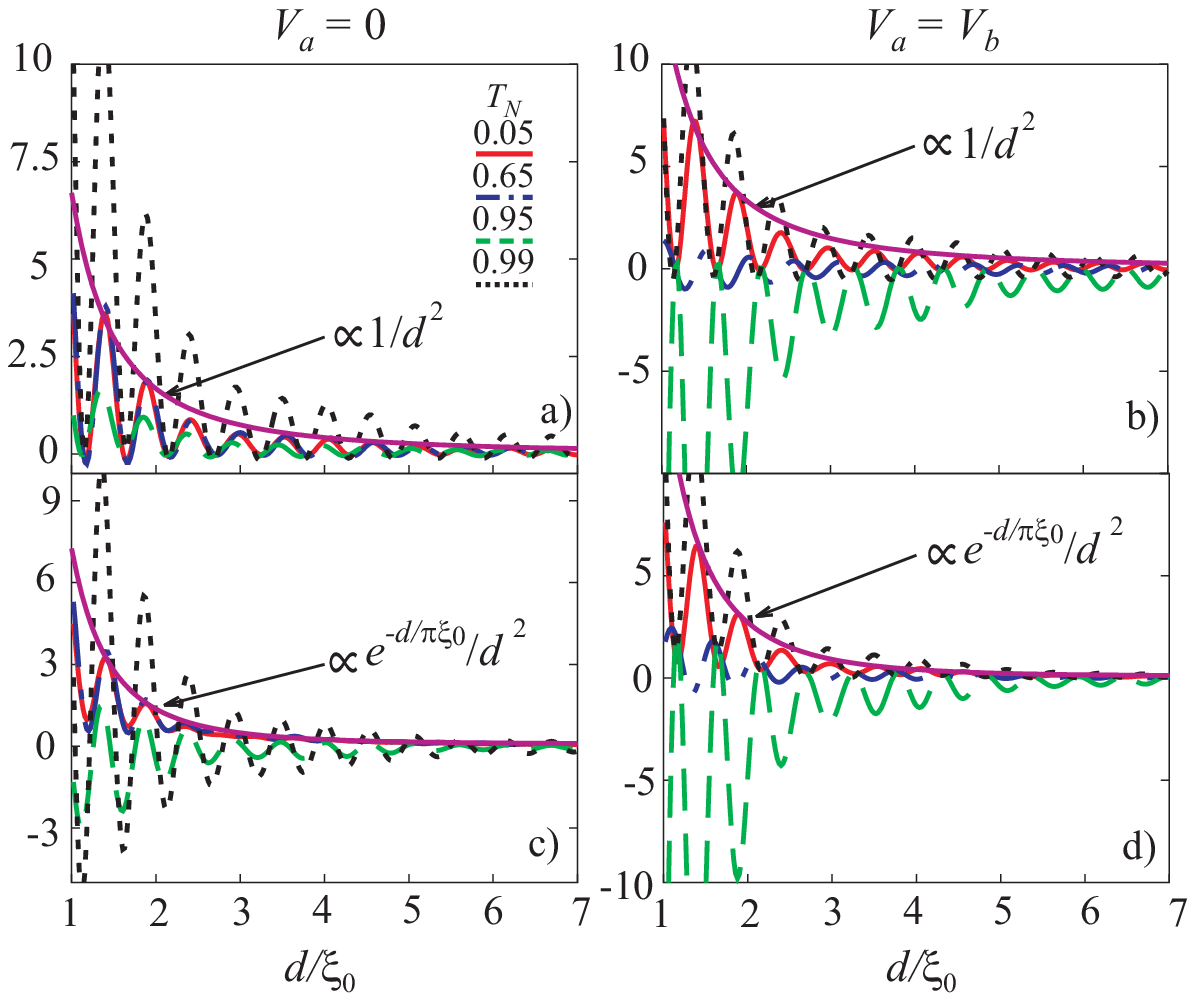}
\caption{Differential shot noise as a function of the distance between the
two leads for different transmission coefficient values $T_N$, a) and b) for
a $d_{x^{2}-y^{2}}$ wave superconductor and c) and d) for a $%
d_{x^{2}-y^{2}}+is$ wave superconductor, with non-symmetric voltages and
symmetric voltages in the first and second column respectively. The
isotropic component for the mixed symmetry is $0.05\ast \Delta _{0}$.}
\label{Fig.7}
\end{figure}

We add an isotropic component to the $d_{x^{2}-y^{2}}$ pair potential $%
\Delta _{\pm}=\Delta _{0}\cos \left( 2(\theta \mp \alpha )\right) +i\Delta
_{s}$ to get the $d_{x^{2}-y^{2}}+is$ symmetry, with $\Delta _{s}=0.05\Delta
_{0}$.

For non-symmetric voltages, in the tunneling limit, we get positive
differential shot noise similarly to those observed with the $%
d_{x^{2}-y^{2}} $ symmetry, see Fig. \ref{Fig.7} c). For intermediate
transparencies, the EC, Q and AR contributions can be more relevant than
CAR, producing an oscillating behavior around zero. For symmetric voltages,
the CAR domains over Q and AR for small transparencies and similarly to the
non-symmetric voltages, see Fig. \ref{Fig.7} d), we appreciate that the
differential noise cross-correlation has been displayed for positive values.

Due to the $\Delta_s$ component an exponential decay is added to the
algebraic decay that we observed with the $d_{x^2-y^2}$ wave
superconductors, such that the behavior is proportional to $e^{-d/\pi\xi_0}/{%
d^2}$.

\begin{figure}[h]
\centering{}\includegraphics[width=7.3cm]{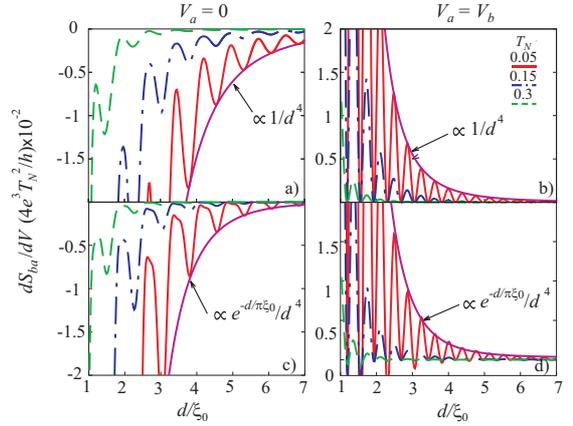}
\caption{Differential shot noise as a function of the distance between the
two leads for different transmission coefficient values $T_N$. a) and b) for
a $d_{xy}$ wave superconductor and c) and d) for a $d_{xy}+is$ wave
superconductor, with non-symmetric voltages and symmetric voltages in the
first and second column respectively. The isotropic component for the mixed
symmetry is $0.05\ast \Delta _{0}$.}
\label{Fig.8}
\end{figure}

When we consider a $d_{xy}$ symmetry the AR are suppressed by the diffraction in the interphase between the lead and superconductor \cite{William_nano}; thus, the processes that contribute to the shot noise are CAR, EC and Q. In NIS (N: normal-metal, I: Insulator, S: Superconductor) plane junctions, when the symmetry of the pair potential is $d_{xy}$, a zero bias conductance peak (ZBCP) appears in the differential conductance due to the induction of states at the interface. The Andreev reflection coefficient is 1 and rapidly decays as $V$ increases. In the differential shot noise the peak is split and at zero bias the shot noise is equal to zero \cite{PhysRevLett.74.3451, 0034-4885-63-10-202, Tanaka2000, Zhu1999}. These two results lead to a Fano factor equal to zero at zero bias for this kind of symmetry. However, in NIS quantum point contact ZBCP does not appear because the Andreev reflections are zero. The wave functions in the channel are a superposition of two plane waves with wave numbers $ky=\pm p/W$. Each wave experiences a pair potential phase $0$ and $\pi $, respectively, and therefore the Andreev reflection coefficient for each wave is out of phase by $\pi$, such that the waves of the reflected holes interfere destructively and the Andreev reflections vanish.

When the leads are connected to non-symmetric voltages and at the tunneling
limit, the EC dominates over CAR, so that the noise cross-correlation
displays negative values. When $T_{N}$ increases, the $Q$ contributions
increase also, causing that the differential noise cross-correlation becomes
more negative. Unlike the $d_{x^{2}-y^{2}}$ symmetry, for $d_{xy}$ symmetry,
we always have domain of negative cross correlations, see Fig. \ref{Fig.8}.

Considering symmetric voltages, we obtain positive values for the shot noise
cross correlations. As the EC contributions to the cross correlations cancel
and the AR are suppressed due to the pair potential symmetry, we only have
the CAR and Q contributions. For low transparencies we appreciate a
dominance of CAR that decreases as $T_{N}$ increases.

Whereas for $d_{xy}$ and non-symmetric voltages the dominance is due to EC,
with an isotropic component the AR occur and contributions due to CAR
increase. This phenomenon can be appreciated in the negative values obtained
in the differential shot noise with an isotropic component, see Fig. \ref%
{Fig.8} a).

For the $d_{xy}$ symmetry we obtain an algebraic decay with respect to the
distance of the leads proportional to $1/{d^{4}}$ (Fig. \ref{Fig.8} a) and
b)), whereas for the $d_{xy}+is$ symmetry, the decay is proportional to $%
e^{-d/{\pi \xi _{s}}}/{d^{4}}$ ( Fig. \ref{Fig.8} c) and d)).

\subsection{Multiband superconductors, $s_{++}$ and $s_{+-}$ symmetries}

The main feature of iron based superconductors is their multiple band
structure near to the Fermi level. In a simplified scheme, the structure is
reduced to two band models, where the symmetry of the pair potential in each
band could be different, and experimental evidence has been favorable to the
$s_{++}$ and $s_{+-}$ symmetries \cite{Kamihara,Golubov,Zhang}.

\begin{figure}[h]
\centering{}\includegraphics[width=7.3cm]{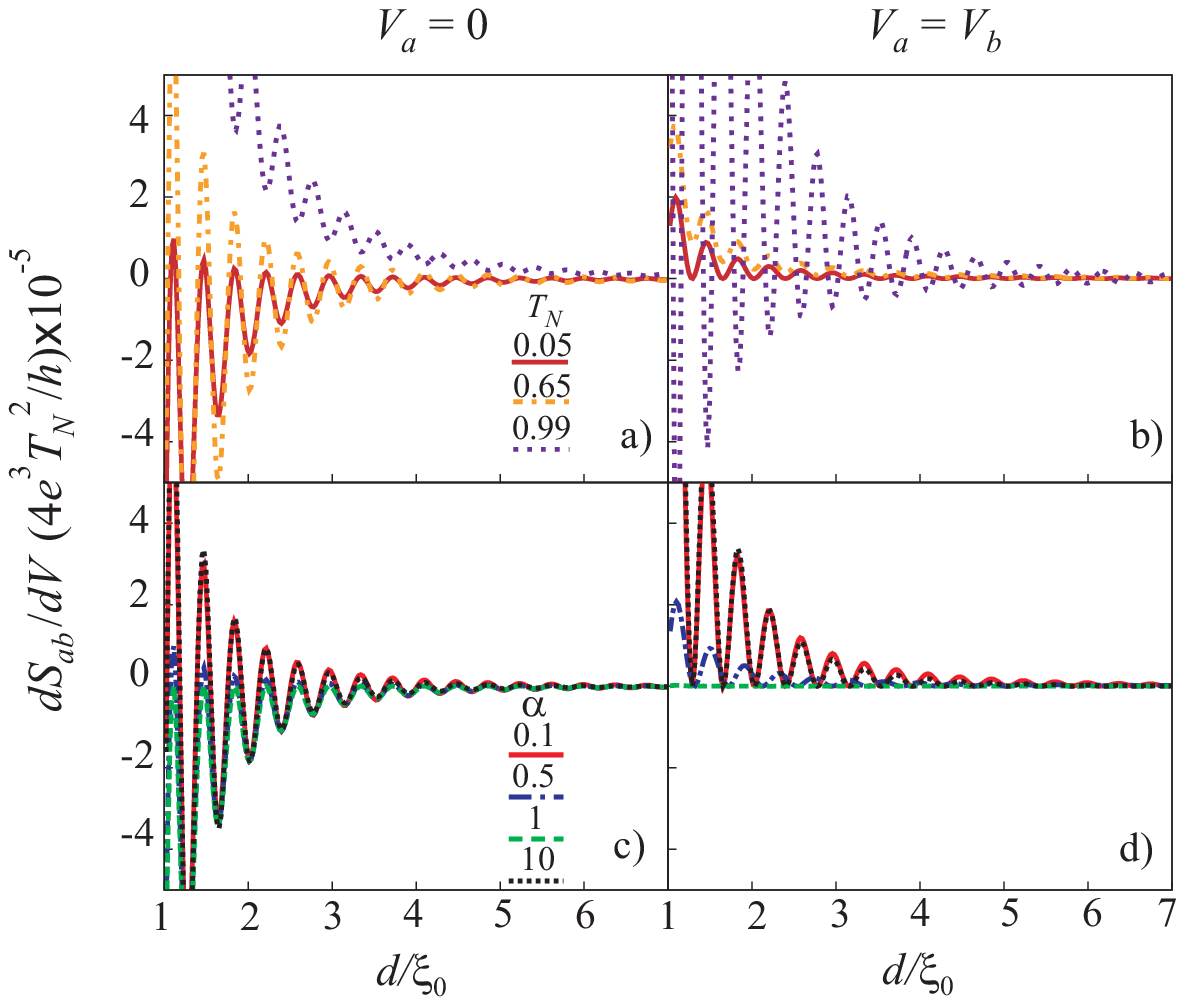}
\caption{Differential shot noise as a function of the distance between the
two leads for a $s_{+-}$ wave superconductor. a) and b) for different
transmission coefficient values $T_N$ and $\protect\alpha =0.5$, c) and d)
for different weight factors $\protect\alpha$ and $T_N=0.05$. With
non-symmetric voltages and symmetric voltages in the first and second column
respectively.}
\label{Fig.9}
\end{figure}

In order to calculate the transport properties for this kind of symmetries,
we find the equilibrium Green functions for two $s$-wave superconductors
with different pair potentials, $\Delta _{1}=0.5\Delta _{0}$ and $\Delta
_{2}=\Delta _{0}$. We use two phase differences between the two gaps, which
for the $s_{++}$ is $0$ and for the $s_{+-}$ is $\pi $. Then by means of a
weight factor ($\alpha $), defined as the ratio of the probability
amplitudes for an incoming electron from one of the two leads to tunnel into
the first or second band, we write the total green function of the system as
\cite{Celis}

\begin{eqnarray}
\hat{g}_{RR}^{r}(E) &=&\sum_{k_{y}}|t(k_{y})|^{2}\left[ \hat{g}_{\Delta
_{1}}^{r}(E,k_{y})\right. \\
&&\left. +\alpha \ \hat{g}_{\Delta _{2}}^{r}(E,k_{y})\right] .  \nonumber
\end{eqnarray}

For the $s_{+-}$ symmetry and non-symmetric voltages, we appreciate
destructive effects over AR due to the phase difference in the pair
potential (Fig. \ref{Fig.9} a) and b) ). However, the contributions due to
AR do not disappear completely because the magnitude of the two gaps are not
equal. At the tunneling limit, the cross-correlations are negative due to
the larger EC contribution. For high transparencies the positive
contributions to the noise cross correlation increase because of non-local
processes of higher order.

When we set symmetric voltages, the EC contributions cancel and the AR are
reduced, the CAR contributes positively to the cross-correlation; thus, the
shot noise is positive.

\begin{figure}[h]
\centering{}\includegraphics[width=7.3cm]{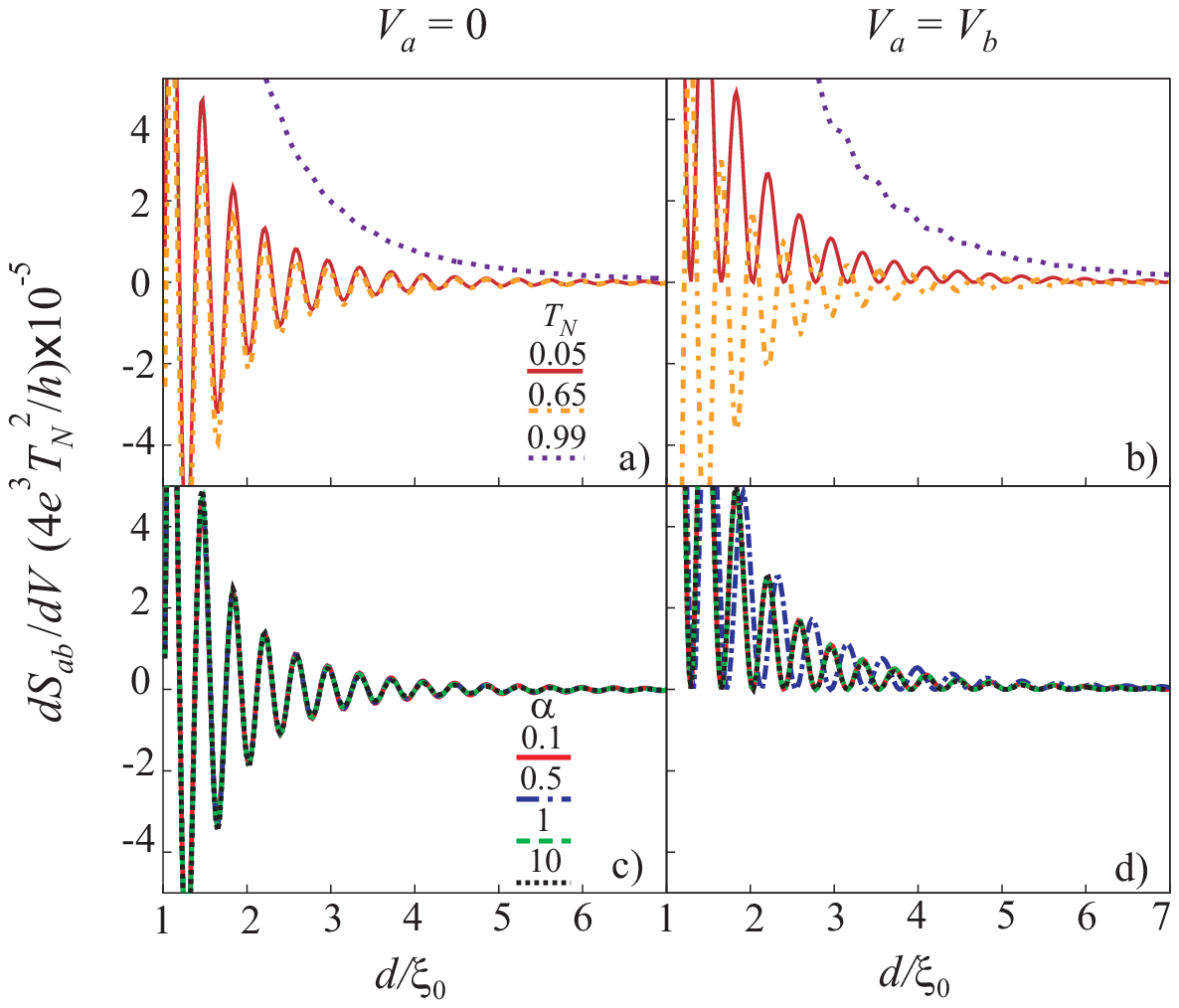}
\caption{Differential shot noise as a function of the distance between the
two leads for a $s_{++}$ wave superconductor. a) and b) for different
transmission coefficient values $T_N$ and $\protect\alpha =0.5$, c) and d)
for different weight factors $\protect\alpha$ and $T_N=0.05$. With
non-symmetric voltages and symmetric voltages in the first and second column
respectively.}
\label{Fig.10}
\end{figure}

The results for the $s_{++}$ symmetry in Fig. \ref{Fig.10} are quite similar
to those obtained for the $s$ symmetry. We also appreciate that the nonlocal
shot noise for symmetric voltages becomes negative for intermediate values
of the transparency according to the behavior observed in the Fig. \ref%
{Fig.4}. This behavior occurs because for this pair potential there is no
phase difference; thus, the contributions of the two bands to the nonlocal
shot noise are added, yielding similar magnitudes and signs as those for $s$
symmetry.

\section{Conclusions}

We have determined analytical equations for shot noise cross correlation in
multi-terminal superconductors from the Green's functions and the Keldysh
formalism. We have considered two leads ($a$ and $b$) and the cases where
the applied voltages are $V_{a}=0$ and $V_{b}=V$ (non-symmetric case), and $%
V_{a}=V_{b}=V$ (symmetric case). We have considered pair potential
symmetries, $s$, $d$, $d+is$ and multiband $s_{+-}$ and $s_{++}$. We found
that when we apply symmetric voltages lower than $|\Delta (\mathbf{k})|$ in
the tunneling limit, positive cross correlations are favored. This result is
obtained due to the fact that the EC, which are the main source of negative
cross correlations, is canceled allowing the CAR to be the dominant
processes.

We calculate the crossed Fano factor of the system for symmetric voltages.
The sign of the Fano factor reveals the sign of the cross correlation
dominance and we find that this factor in the tunneling limit is analogous
to the CPS efficiency of the device.

In particular, we observe that when symmetric voltages are applied to a $%
d_{xy}$-wave superconductor, and in the tunneling limit, the CAR become the
dominant processes making the device to exhibits good efficiency even for
voltages higher than $\Delta _{0}$. By on the other side, for non symmetric
voltages EC dominates over CAR. For $d_{x^{2}-y^{2}}+is$ and $d_{xy}+is$
symmetries, the isotropic component causes negative contributions reflected
in negative values for the cross correlation.

In general the shot noise cross correlation between two leads for $d$-wave
superconductors exhibits algebraic decay with the increase of the distance
between the leads, in contrast to the exponential behavior typically
observed for isotropic superconductors. These properties would allow the
development of devices that are capable of detecting positive cross
correlation at distances several times larger than the characteristic
coherence length with good efficiency.

To prove the entanglement, it is still needed to show Bells inequality by
means of the coherence and spin correlation that could be accomplished by
means of ferromagnetic leads.


The authors have received support from the COLCIENCIAS, project No.
110165843163.

\appendix

\section{Modeling the system}

\label{Sec:Modeling the system}

From the superconductor surface Green functions in momentum representation ($%
\hat{g}_{S}^{r}(E,ky)$), we calculate the local $\hat{g}_{S,aa(bb)}(E)$ and
the non-local $\hat{g}_{S,ab(ba)}(E)$ equilibrium Green functions \cite%
{Williamdos}

\begin{eqnarray}
\hat{g}_{S,bb(aa)}^{r}(E)\!\! &=&\!\!\sum_{k_{y}}|f(k_{y})|^{2}\hat{g}%
_{S}^{r}(E,k_{y}),  \nonumber \\
\hat{g}_{S,ba(ab)}^{r}(E)\!\! &=&\!\!\sum_{k_{y}}|f(k_{y})|^{2}\hat{g}%
_{S}^{r}(E,k_{y})  \label{non-local_green} \\
&&e^{-(+)ik_{y}d/\xi },  \nonumber
\end{eqnarray}%
where $f\left( k_{y}\right) $ is the weighting factor and is proportional to
the perpendicular wave vector $k_{xF}$ \cite{William_nano}, $\xi $ is the
superconductor coherence longitude defined as $\xi (E,\theta )=\xi _{0}/Re[%
\sqrt{1-E^{2}/\Delta (\theta )}]$ and $\xi _{0}=\hbar \nu _{F}/(\pi \Delta
_{0})$ the BCS coherence longitude.

We write the retarded Green function of the uncoupled superconductor $\check{%
g}^r_{RR}\left( E\right)$ as

\begin{equation}
\check{g}_{S}^{r}\left( E\right) =\left(
\begin{array}{cc}
\hat{g}_{S,aa}^{r}\left( E\right) & \hat{g}_{S,ab}^{r}\left( E\right) \\
\hat{g}_{S,ba}^{r}\left( E\right) & \hat{g}_{S,bb}^{r}\left( E\right)%
\end{array}%
\right) ,
\end{equation}%
where the symbol $\hat{}$ denotes a $2\times 2$ matrix in Nambu space,
whereas the symbol $\check{}$ denotes a $4\times 4$ matrix in the Nambu
electrodes space.

We obtain the non-equilibrium Green functions of the coupled system $\check{G%
}_{ij,\beta \beta ^{\prime }}^{+-(-+)}(E)$ and the perturbed Green Function $%
\check{G}_{ij,\beta \beta ^{\prime }}^{r(a)}(E)$ solving the Dyson equation
for two leads, where $\beta $ and $\beta ^{\prime }$ denote the $a$ or $b$
lead respectively, and $i$ and $j$ denote the $L$ or $R$ region respectively.

\begin{eqnarray}
\check{G}^{+-(-+)}\!(E)\! &=&\![\check{I}\!+\!\check{G}^{r}\!(E)\check{p}]%
\check{g}^{+-(-+)}\!(E)  \label{Gmn} \\
&&\lbrack \check{I}\!+\!\check{p}\check{G}^{a}\!(E)],  \nonumber \\
\check{G}^{r(a)}(E) &=&\check{g}^{r(a)}(E)+\check{g}^{r(a)}(E)\check{p}
\label{Gra} \\
&&\check{G}^{r(a)}(E),  \nonumber
\end{eqnarray}%
here $t$ is the hopping parameter related to the transmission of particles
from the left side to the right side \cite{William_nano}, $+(-)$ are the two
branches of the Keldysh space. $\check{g}^{+-(-+)}$ is the non-equilibrium
Green function (leads or superconductor) in Keldysh space without coupling
\cite{Williamdos}.

The electrical current in the $\beta $ lead is
\begin{eqnarray}
I_{\beta } &=&\frac{et^{2}}{h}\int_{-\infty }^{\infty }\!dE\mathrm{Tr}\{\!%
\check{N}_{\beta }\left( \check{g}_{LL}^{+-}(E)\check{\sigma}_{z}\right.
\!\!\! \\
&&\left. \check{G}_{RR}^{-+}(E)-\check{g}_{LL}^{-+}(E)\check{\sigma}_{z}%
\check{G}_{RR}^{+-}(E)\right) \},  \nonumber
\end{eqnarray}%
with, $\check{N}_{a},\check{N}_{b},$ $\check{\sigma}_{z}$ given by the Eq. %
\ref{MAtrixN}. To calculate the noise cross-correlation we define the
spectral density of the electrical current fluctuations between the
electrodes $\beta $ and $\beta ^{\prime }$ as

\begin{eqnarray}
S_{\beta \beta ^{\prime }}\left( \omega \right) &=&\hbar \!\int \!d(\tau
^{\prime })e^{i\omega (\tau ^{\prime })}\!  \nonumber \\
&&\left[ \left\langle \delta \hat{I}_{\beta }(\tau ^{\prime })\delta \hat{I}%
_{\beta ^{\prime }}(\tau )\right\rangle +\right. \\
&&\left. \left\langle \delta \hat{I}_{\beta ^{\prime }}(\tau ^{\prime
})\delta \hat{I}_{\beta }(\tau )\right\rangle \right] ,  \nonumber
\end{eqnarray}%
where $\delta \hat{I}_{\beta }(\tau )$ is the deviation of the electrical
current regarding to its mean value, $\delta \hat{I}_{\beta }(\tau )=\hat{I}%
_{\beta }(\tau )-\langle \hat{I}_{\beta }(\tau )\rangle $.

\begin{eqnarray}
\left\langle \delta \hat{I}_{\beta }(\tau ^{\prime })\delta \hat{I}_{\beta
^{\prime }}(\tau )\right\rangle &=&2\left( \frac{e}{\hbar }\right) ^{2}%
\mathrm{Tr}\left[ \check{N}_{\beta }\check{\sigma}_{z}\check{p}\right.
\nonumber \\
&&\check{G}_{LL}^{+-}(\tau ,\tau ^{\prime })\sigma _{z}\check{p} \\
&&\check{G}_{LL}^{-+}(\tau ^{\prime },\tau )-\check{N}_{\beta }\check{\sigma}%
_{z}\check{p}  \nonumber \\
&&\check{G}_{LR}^{+-}(\tau ,\tau ^{\prime })\sigma _{z}\check{p}  \nonumber
\\
&&\left. \check{G}_{LR}^{-+}(\tau ^{\prime },\tau )\right] ,  \nonumber
\end{eqnarray}%
with $\check{p}=t\ast \check{\sigma}_{z}.$ We write the noise
cross-correlation as

\begin{eqnarray}
S_{\beta \beta ^{\prime }}\left( \omega \right) &=&\frac{2e^{2}t^{4}}{h}\int
dE\left[ K_{\beta \beta ^{\prime }}(E,\varepsilon )\right.  \nonumber \\
&&\left. +K_{\beta ^{\prime }\beta }(\varepsilon ,E)\right] ,
\label{ruido2c} \\
\varepsilon &=&E+\hbar \omega ,  \nonumber
\end{eqnarray}%
where the kernel $K_{\beta \beta ^{\prime }}\left( E,\varepsilon \right) $
is given by

\begin{eqnarray}
K_{\beta \beta ^{\prime }}\left( E,\varepsilon \right) &=&Tr\!\left[ \check{N%
}_{\beta }\check{G}_{LL}^{+-}\left( E\right) \check{N}_{\beta ^{\prime
}}\right.  \label{K_Gmn} \\
&&\check{G}_{RR}^{-+}\left( \varepsilon \right) -\check{N}_{\beta }\check{G}%
_{LR}^{+-}\left( E\right)  \nonumber \\
&&\left. \check{N}_{\beta ^{\prime }}\check{G}_{LR}^{-+}\left( \varepsilon
\right) \right] .  \nonumber
\end{eqnarray}

From the non-equilibrium Green function obtained by solving the Dyson
equation Eq. (\ref{Gra}) we calculate the kernels $K_{\beta \beta ^{\prime
}} $ Eq. (\ref{K_Gmn}) and the noise cross-correlations, Eq. (\ref{ruido2c}).

\bibliographystyle{unsrt}
\bibliography{JACelisGil2}

\end{document}